\documentclass[11pt]{amsart}

\usepackage[bookmarksnumbered,colorlinks,bookmarks,citecolor=blue,linkcolor=blue,urlcolor=blue,breaklinks,linktocpage]{hyperref}
\usepackage[all]{hypcap}

\usepackage{amsmath}
\usepackage{graphicx}
\usepackage{amssymb}
\usepackage{subfigure}
\usepackage{url}

\title[Benford \&  Music ]{Emergence of Benford's Law in Music}

\author{Azar Khosravani}
\address{Department of Science and Mathematics, Columbia College Chicago,
	Chicago, IL 60605}
\email{akhosravani@colum.edu}

\author{Constantin Rasinariu}
\address{Department of Physics, Loyola University Chicago,
	Chicago, IL 60660}
\email{crasinariu@luc.edu}



\begin{document}

\maketitle

\begin{abstract}
	We analyzed a large selection of classical musical pieces composed by Bach, Beethoven, Mozart, Schubert and Tchaikovsky, and found a surprising connection with mathematics. For each composer, we extracted the time intervals each note was played in each piece and found that the corresponding data sets are Benford distributed.  Remarkably, the logarithmic distribution is present not only for the leading digits, but also for all digits.
\end{abstract}
	
\section{Introduction}

What does the Moonlight Sonata by Beethoven have in common with the Swan Lake ballet by Tchaikovsky? They both exhibit Benford distributed time intervals for their constituent musical notes. This result is not unique. We analyzed hundreds of musical pieces composed by Bach, Beethoven, Mozart, Schubert, and Tchaikovsky and found that in each case, the note durations were Benford distributed. 

Our data  consists of a selection of MIDI files downloaded from the music archive \url{http://www.kunstderfuge.com}, which is a major resource housing thousands of music files. We chose a collection of sonatas, concertos, etc., for a total of $521$ files. 
Depending on the structure of each musical piece, the piece may be spread over several files. For instance, Tchaikovsky's Swan Lake has 4 acts with each act broken to parts for a total of $43$ files. We used Mathematica \cite{Wolfram-11.3} to obtain the time duration each note was played in a given file. In our analysis we ignored the dynamics, thus the quieter parts were given the same weight as the louder ones. Data was compiled into tables, which were analyzed for their digit distributions. With no exception, we observed the emergence of Benford's law across the works of each of the composers we studied. 

This paper is structured as follows. First, we present a short introduction to Benford's law.  Then, we present our digit distribution analysis for the time duration tables for all classical pieces mentioned above.        
We used a Quantile-Quantile (Q-Q) representation in which the experimental data sets were plotted against the theoretical Benford distribution and found a remarkable close agreement.

\section{Benford's Law}
\label{sec:Ben}

\begin{figure}[htb]
	\centering
	\includegraphics[width=0.5\linewidth]{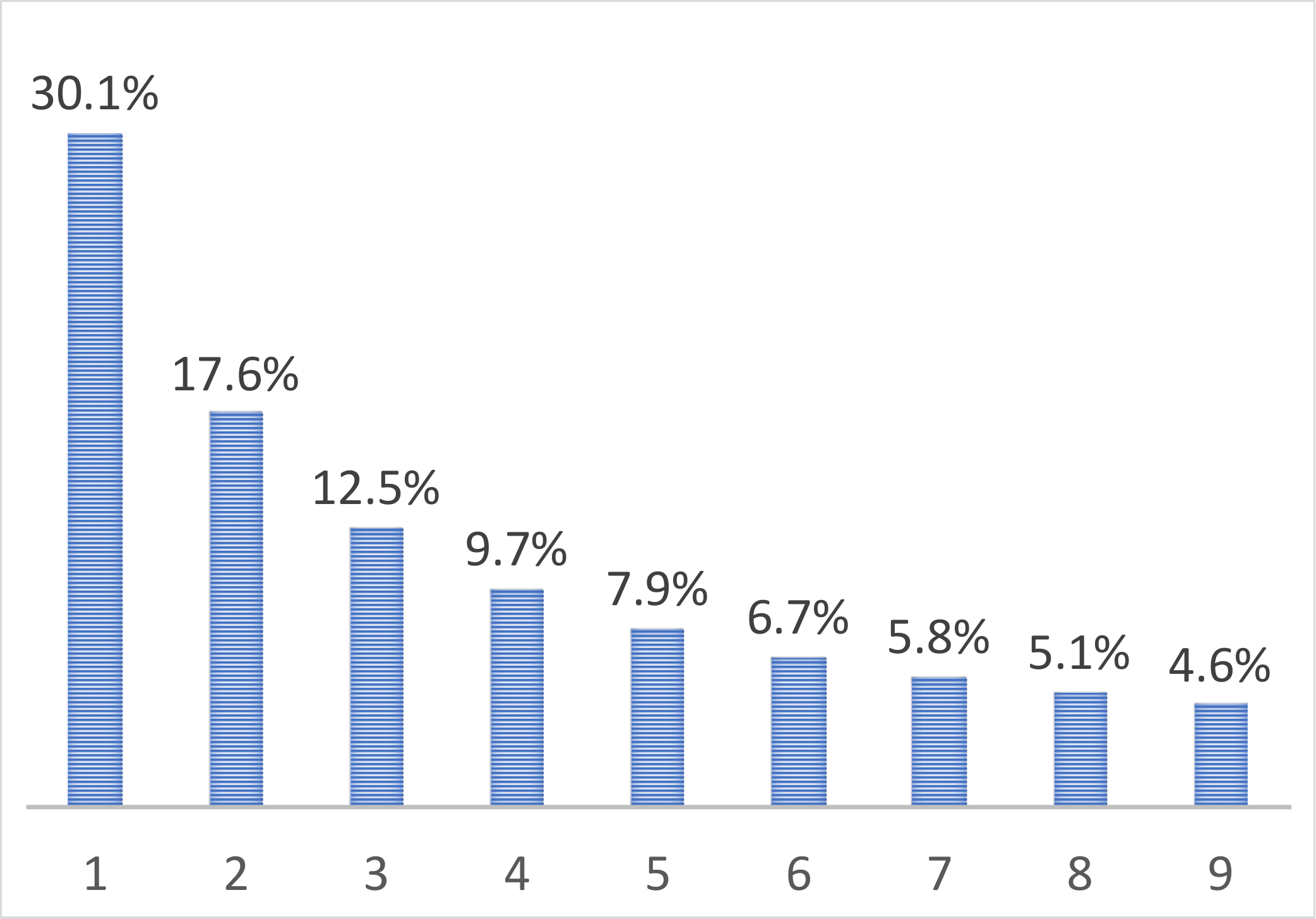}
	\caption{Benford's Law for the first digit}
	\label{fig:fig1}
\end{figure}

Benford's law comes from the empirical observation that in many data sets the leading digits of numbers are more likely to be small than large, for instance, $1$ is more likely to occur as the leading digit than $2$, which in turn is more likely the first digit than $3$, etc. This observation was first published by Newcomb in 1881 \cite{Newcomb-1881}, and given  experimental support in 1938 by Benford \cite{Benford-1938} who analyzed over $20,000$ numbers collected from naturally occurring data sets such as the area of the riverbeds, atomic weights of elements, etc.  Explicitly, he showed that the probability of $d$ being the first digit is 
\begin{equation}
\label{eq:Ben}
P_d = \log_{10}\left(1 + \frac 1d \right),\quad d=1,2,\ldots, 9
\end{equation}
which came to be known as Benford's law. The first digit probabilities are illustrated in Fig. (\ref{fig:fig1}).  
Similarly, there are logarithmic expressions for the probabilities of the second, third and other digits. For instance, the probability of a number having its first digit $d_1$ and second digit $d_2$ is
\begin{equation}
\label{eq:Ben2}
P_{d_1d_2} = \log_{10}\left( 1+ \frac{1}{d_1d_2} \right) .
\end{equation}
Thus, in a Benford distributed data set, the probability of a number having its first digit equal to 2 and the second digit equal to 6 is $P_{26} = \log_{10}\left( 1+ \frac{1}{26} \right) \simeq 0.0164$.

In general, a set $\{x_n\}$ of real positive numbers is Benford \cite{Berger-2011} if
\begin{equation}
\label{eq:CDF}
\lim_{N\to\infty}\frac{\#\{1\leq n \leq N:S(x_n) \leq t\}}{N} = \log t ,\quad 
\textrm{for all } t\in [1,10)
\end{equation}
where
\begin{equation}
\label{eq:Sig}
S(x)=10^{\log_{10}x - \log_{10}\lfloor x \rfloor}      ,\quad x>0~
\end{equation}
is the significand function $S:\mathbb R^{\,+} \to [1,10)$. In the above definition $\lfloor x \rfloor$ denotes the floor function. The significand function simply gives the first part of the scientific notation of any number. For example, the significand of $143$ is $S(143) = 1.43$.  

Benford distributed sequences have several intriguing characteristics. First, they are scale invariant. That is, if one multiplies all the elements of the sequence by a scalar, the resulting sequence will be Benford distributed \cite{Pinkham-1961,Raimi-1976}. Second, Benford sequences are base invariant \cite{Hill-1995}. This means that there is nothing special about base $10$. For a general base $b$ the first digit formula reads 
\begin{equation}
\label{eq:Benb}
P_d = \log_{b}\left(1 + \frac 1d \right),\quad d=1,2,\ldots, b-1.
\end{equation}
The third property concerns the uniform distribution \cite{Diaconis-1977} of the logarithm base $b$ of Benford sequences. Namely, a sequence $\{x_n\}$ is Benford if and only if $\{\log_{b}x_n\}$ is uniformly distributed mod 1. A fourth property concerns the sum invariance \cite{Nigrini-1992,Allaart-1997}. Let  $\{S(x_n)\}$ be the sequence of significands, as defined by Eq. (\ref{eq:Sig}), of a Benford distributed sequence $\{x_n\}$. Define the sum of all significands of numbers starting with $i$ as $S_i$. Sum invariance in the first digit means that $S_i = S_j$ for all $i,j = 1,2,\cdots 9$. In other words, the sum of all significands of numbers starting with $1$ is equal to the sum of all significands of numbers starting with $2$, and so on. This can be generalized to more digits. For example, in the case of the first two digits, the sum invariance implies that the sum of all numbers with significands starting with $10$ through $99$ are equal. 

The current research reaffirms the ubiquity of Benford's law in many collections of numerical data. For a large list of applications including fraud detection in financial data \cite{Nigrini-1992,Nigrini-2012}, survival distributions \cite{Leemis-2000}, and distances from Earth to stars \cite{Alexopoulus-2014}, see \cite{BOL-2011}. In this paper we would like to add one more instance of emergence of Benford's law: music.

\section{Data Extraction and Analysis}
\label{sec:MIDI}

We chose a collection of sonatas, concertos, etc., for a total of 521 MIDI files and, using Mathematica, extracted the time duration each note was played in a given musical piece. For example, for Sonata no. 14 in C\# minor "Quasi una fantasia", Opus 27, No. 2, also known as the Moonlight Sonata, by Beethoven, we have obtained the data summarized in Table \ref{Moon}. Each pair of cells contains the note and its corresponding cumulative play time  in seconds.

\begin{figure}[tbh]
	\centering
	\includegraphics[width=0.5\linewidth]{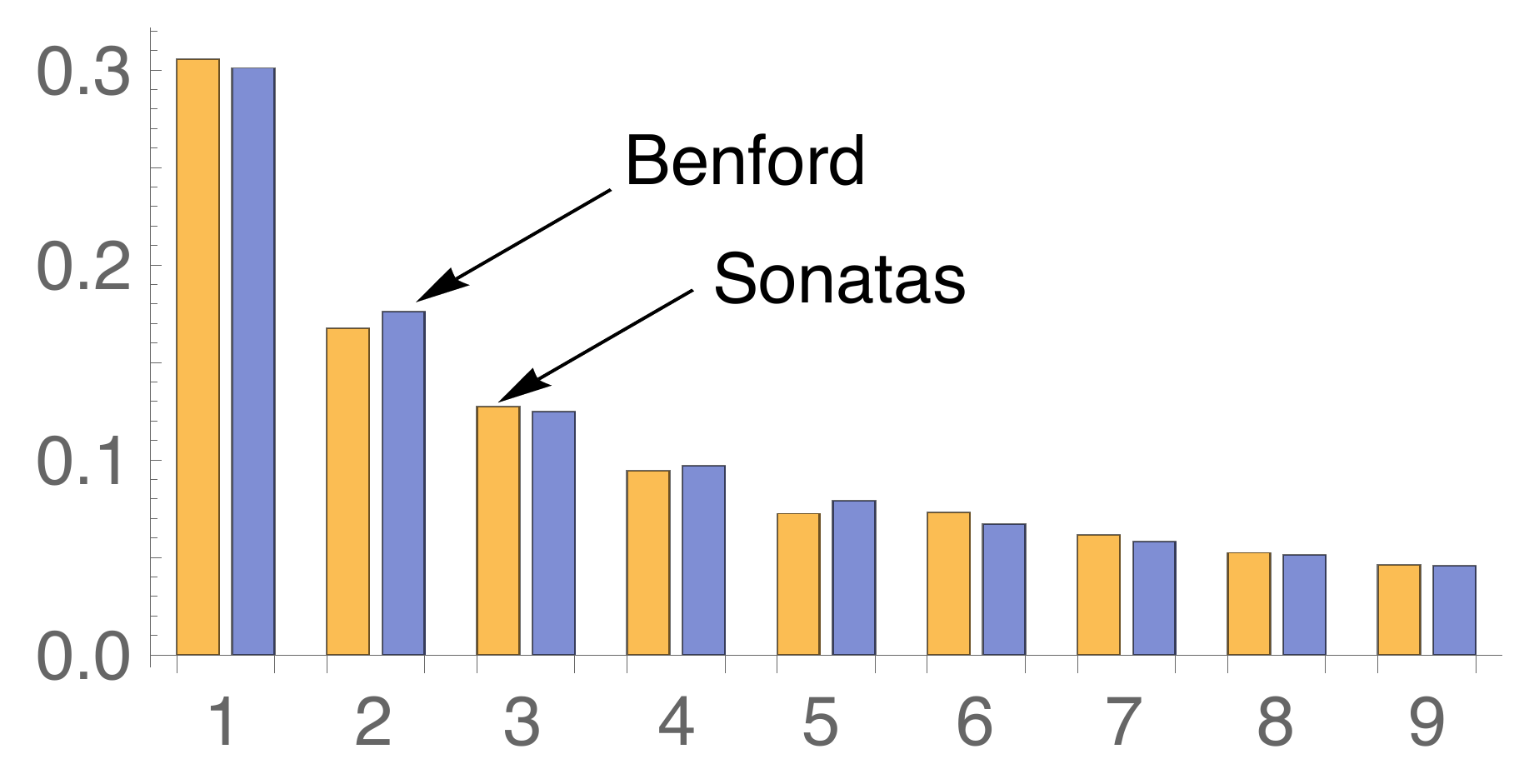}
	\caption{Comparison of the first digit frequencies in Beethoven's sonatas note durations versus Benford distribution.}
	\label{fig:beeth-1d}
\end{figure}

\begin{table}[htb]
\begin{center}
{\tiny
\noindent 
\begin{tabular}{|c|c|c|c|c|c|c|c|c|c|}
	\hline
	F1 & F\#1  & G1 & G\#1 & A1 & A\#1 & B1 & C2 & C\#2 & D2 \\
	\hline 
	12.5704 & 60.0901 & 13.192 & 206.461 & 25.9161 & 3.97648 &53.3207 & 37.2194 & 266.582  & 20.9309 \\
	\hline \hline
	 D\#2 & E2 & F2 & F\#2 & G2 & G\#2 & A2 &A\#2 & B2 & C3\\
	\hline 
	46.4842 & 48.0922 & 24.4303& 119.923 & 48.1226 &  496.196 & 49.56 & 40.6535 & 75.1769 & 75.7273 \\
	\hline \hline
	 C\#3  & D3 & D\#3 & E3 & F3 &F\#3 & G3 &G\#3 & A3 & A\#3\\
	\hline 
	288.172 & 35.8046 & 162.87& 114.506 & 77.1172 & 163.914 & 33.9666 & 340.613 & 66.8618 & 54.1357 \\
		\hline \hline
	B3  & C4 & C\#4 & D4 & D\#4 & E4 & F4 &F\#4 & G4 & G\#4\\
	\hline 
	67.6126 & 123.1 & 316.731 & 23.6688 &176.852 & 152.541 & 118.636 & 211.026 & 67.5102 & 264.642 \\
		\hline \hline
	 A4  & A\#4 & B4 & C5 & C\#5 & D5 & D\#5 &E5 & F5 & F\#5\\
	\hline 
	93.9875 & 73.5818 & 96.3341 & 107.068 & 244.132 & 30.3397 & 148.947 & 74.9435 & 47.7797 & 60.1995 \\
		\hline \hline
	 G5  & G\#5 & A5 & A\#5 &B5 & C6 & C\#6 &D6 & D\#6 & E6\\
	\hline 
	23.6263 & 109.545 & 31.6656 & 19.3303 & 30.5796 &8.1659 & 29.3945 & 0.666666 & 8.89245 & 11.7557 \\
	\hline 
\end{tabular} 
}
\end{center}
\label{Moon}
\caption{Cumulative times for all 60 notes  played in Moonlight sonata.}
\end{table}
For each of the $32$ Beethoven's piano sonatas we constructed similar data sets, and formed the data set $\mathcal{A}$ comprised of the union of all the time durations. The numeric set $\mathcal{A}$ has $2043$ time duration values, which corresponds to $32$ (sonatas) $\times~ 88$ (the number of piano keys) minus the total number of notes in all sonatas that were not played. Next, we extracted the first digit of elements of $\mathcal{A}$, obtaining the following set 
\[
\tilde{\mathcal{A}} = \{1, 6, 1, 2, 2, 3, 5, \cdots [2032] \cdots 9, 8, 6, 2 \}
\]
For brevity, in $\tilde{\mathcal{A}} $ we have shown the first elements of the Moonlight sonata as above, omitted the following 2032 values, and showed the first digit values corresponding to notes A\#6, B6, C7, and D\#7, which are the four rightmost keys on the piano with nonzero play time in Sonata 32. 
Table \ref{Moon-d} contains the frequencies of the first digits 1 through 9 in  $\tilde{\mathcal{A}} $ versus the expected frequencies given by the Benford distribution. The corresponding histograms are shown in Figure \ref{fig:beeth-1d}.

\begin{table}[htb]
	\begin{center}
		{\tiny
			\noindent 
\begin{tabular}{|c|c|c|c|c|}
	\hline 
	Digit & Bin counts in $\tilde{\mathcal{A}}$ & Relative frequency data & Benford & Relative error \\ 
	\hline 
	1 & 624 & 0.305433 & 0.301030 & 0.01463 \\ 
	\hline 
	2 & 342 & 0.167401 & 0.176091 & 0.04935 \\ 
	\hline 
	3 & 260 & 0.127264 & 0.124939 & 0.01861 \\ 
	\hline 
	4 & 193 & 0.0944689 & 0.0969100  & 0.02519 \\ 
	\hline 
	5 & 148 & 0.0724425 & 0.0791812 & 0.08511 \\ 
	\hline 
	6 & 149 & 0.0729320 & 0.0669468 & 0.08940 \\ 
	\hline 
	7 & 126 & 0.0616740 & 0.0579919 & 0.06349 \\ 
	\hline 
	8 & 107 & 0.0523740 & 0.0511525 & 0.02388 \\ 
	\hline 
	9 & 94 & 0.0460108 & 0.0457575 & 0.00554 \\ 
	\hline 
\end{tabular}} 
\end{center}
\label{Moon-d}
\caption{Numerical values extracted from all $32$ Beethoven sonatas vs. Benford distribution values.}
\end{table}
\noindent

The fact that the extracted data is logarithmic distributed in the first digit, does not imply that the data set is Benford distributed. There are examples in literature \cite{Khosravani-2013}, of data sets that are logarithmic distributed in the first digit but not in the following ones. However, to show Benford distribution it suffices to show that Eq. \ref{eq:CDF} holds. A commonly used tool to compare the empirical data sets with a given distribution is the Quantile-Quantile (Q-Q) plot. In our Q-Q plots the empirical data is arranged on the horizontal axis and the theoretical Benford distribution on the vertical axis. 

Applying the significand function defined in Eq. (\ref{eq:Sig}) on the set $\mathcal{A}$, we find the significands of all time durations, which maps the data values into $[1,10)$. Sorting the resulting list yields the empirical quantiles. Using (\ref{eq:CDF}), the $k$-th quantile for the theoretical Benford distribution is given by $10^{k/m}$, where $1 \le k \le m$, and $m$ is the desired number of quantiles. For $m=50$, we get the following Q-Q plot for the $32$ Beethoven sonatas.

\begin{figure}[tbh]
	\centering
	\includegraphics[width=0.45\linewidth]{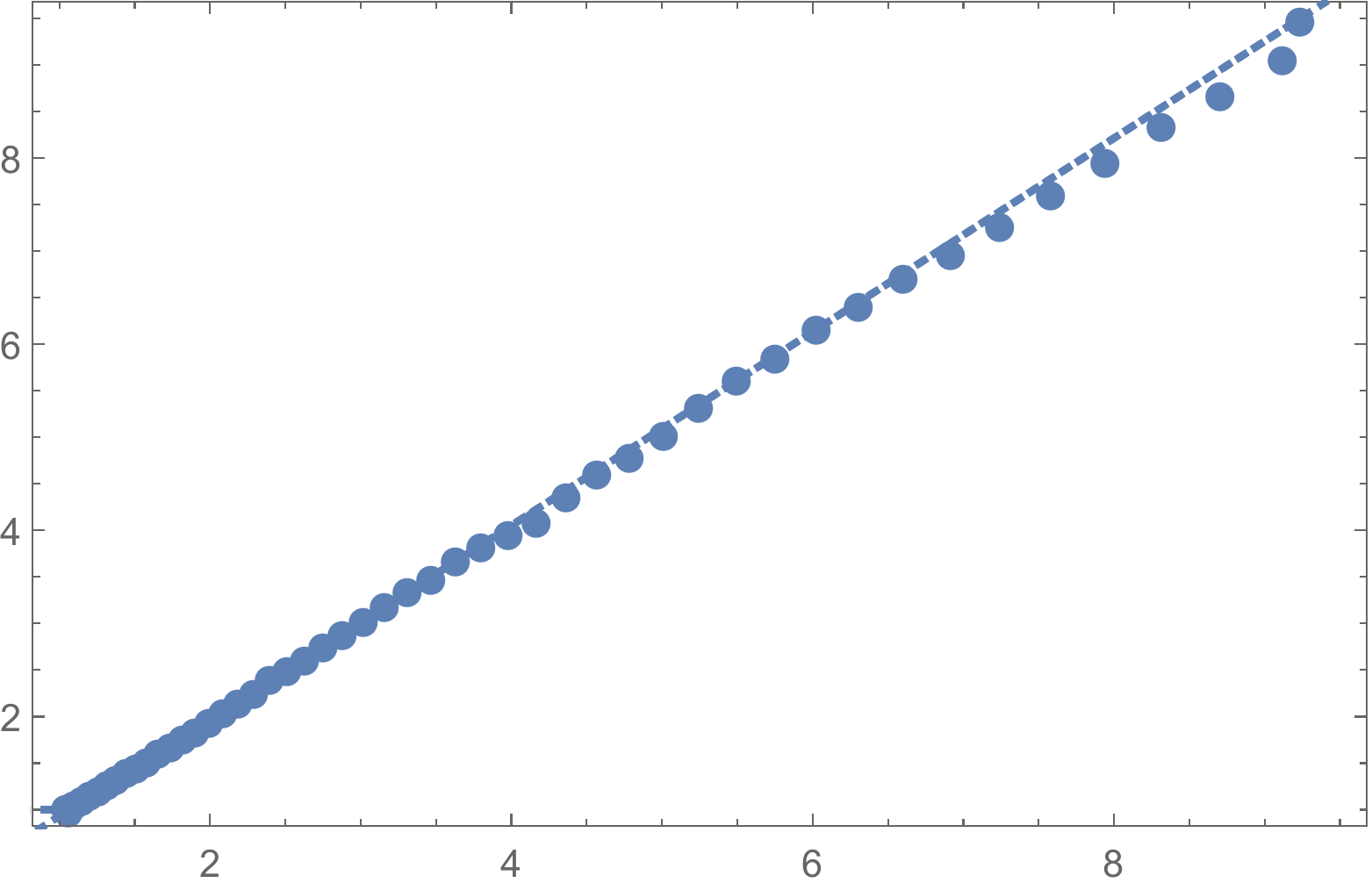}
	\caption{Q-Q plot comparing the theoretical Benford and experimental for the $32$ Beethoven sonatas.}
	\label{fig:q-q-beeth-blue}
\end{figure}

\noindent
Looking at the Figure \ref{fig:q-q-beeth-blue}, one can see that the Q-Q plot points are more concentrated for the lower digits, as expected from the Benford distribution. The linearity of the plot confirms the goodness of fit of the empirical data in all digits.

\begin{figure}[tbh]
	\centering
	\includegraphics[width=0.45\linewidth]{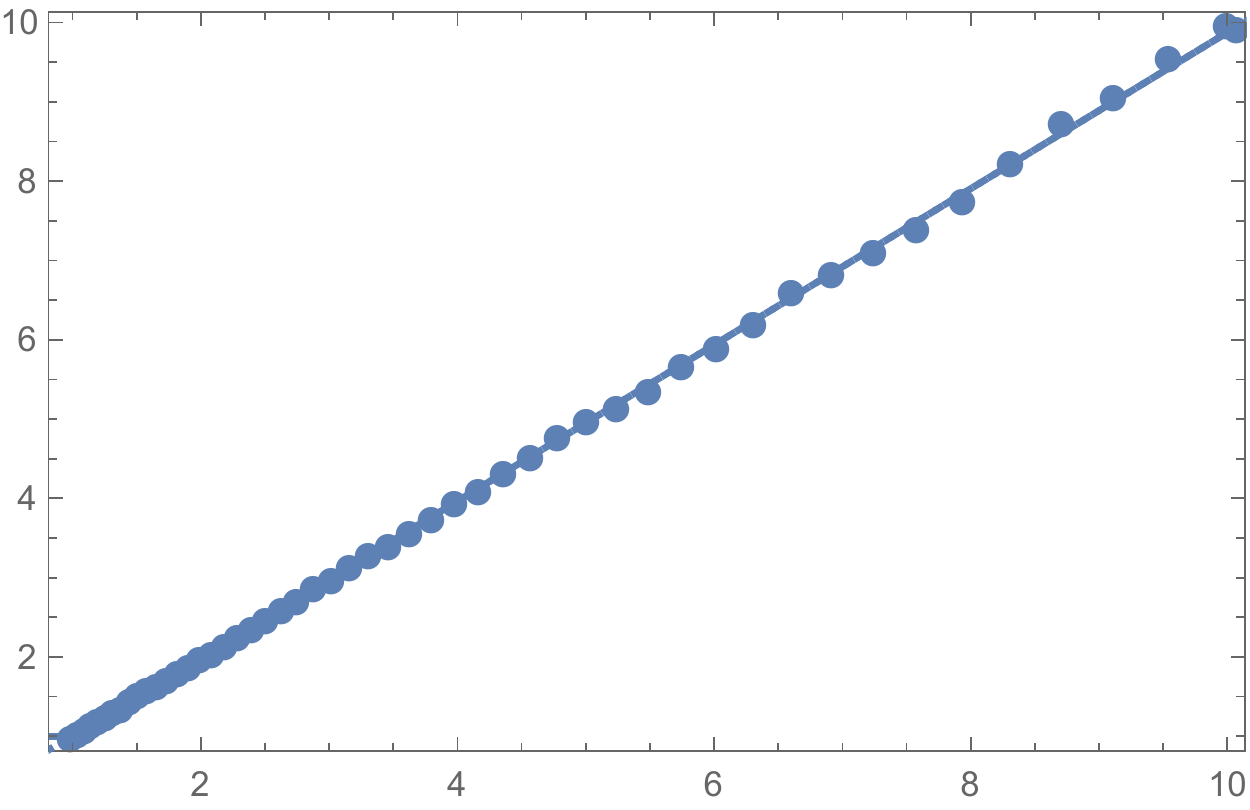}
	\caption{Q-Q plot comparing the theoretical Benford and experimental for the Swan Lake ballet.}
	\label{fig:q-q-tchai50}
\end{figure}

To determine whether similar patterns hold for other composers, we analyzed Tchaikovsky's Swan Lake ballet, which has 4 acts with each act broken to parts for a total $43$ files. As before, we extracted time durations for each note, for each of the $43$ MIDI files, and constructed the data set $\mathcal{A}$. In this case, $\mathcal{A}$ has $2350$ values, corresponding to the non-zero play time notes. Performing a similar analysis on the Swan Lake ballet we get the  Q-Q plot shown in Figure \ref{fig:q-q-tchai50}.

Both Q-Q plots suggest that the data sets obtained from Beethoven's sonatas and Tchaikovsky's Swan Lake are Benford distributed. 

\begin{figure}[htb]
	\label{fig:all}
	\centering
	\subfigure[Bach]{
		\includegraphics[width=0.4\textwidth]{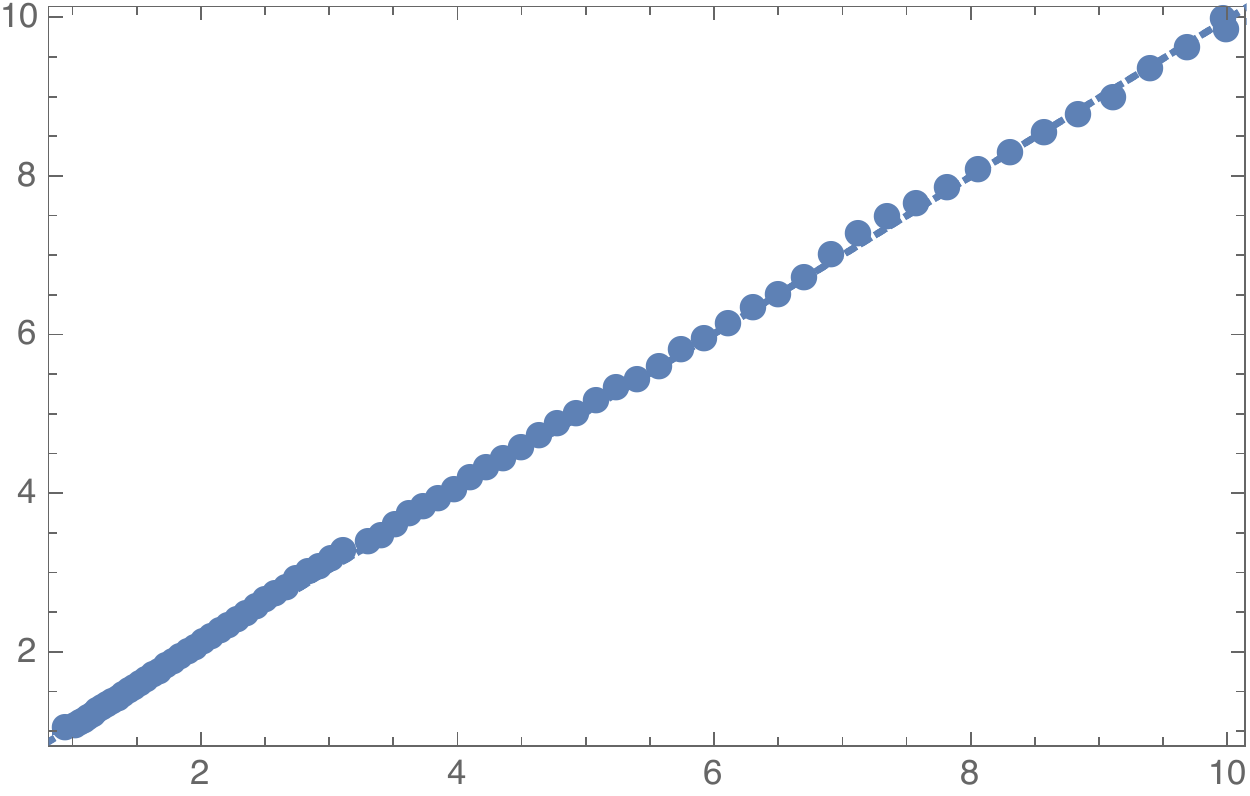}}
	\quad
	\subfigure[Mozart]{
		\includegraphics[width=0.4\textwidth]{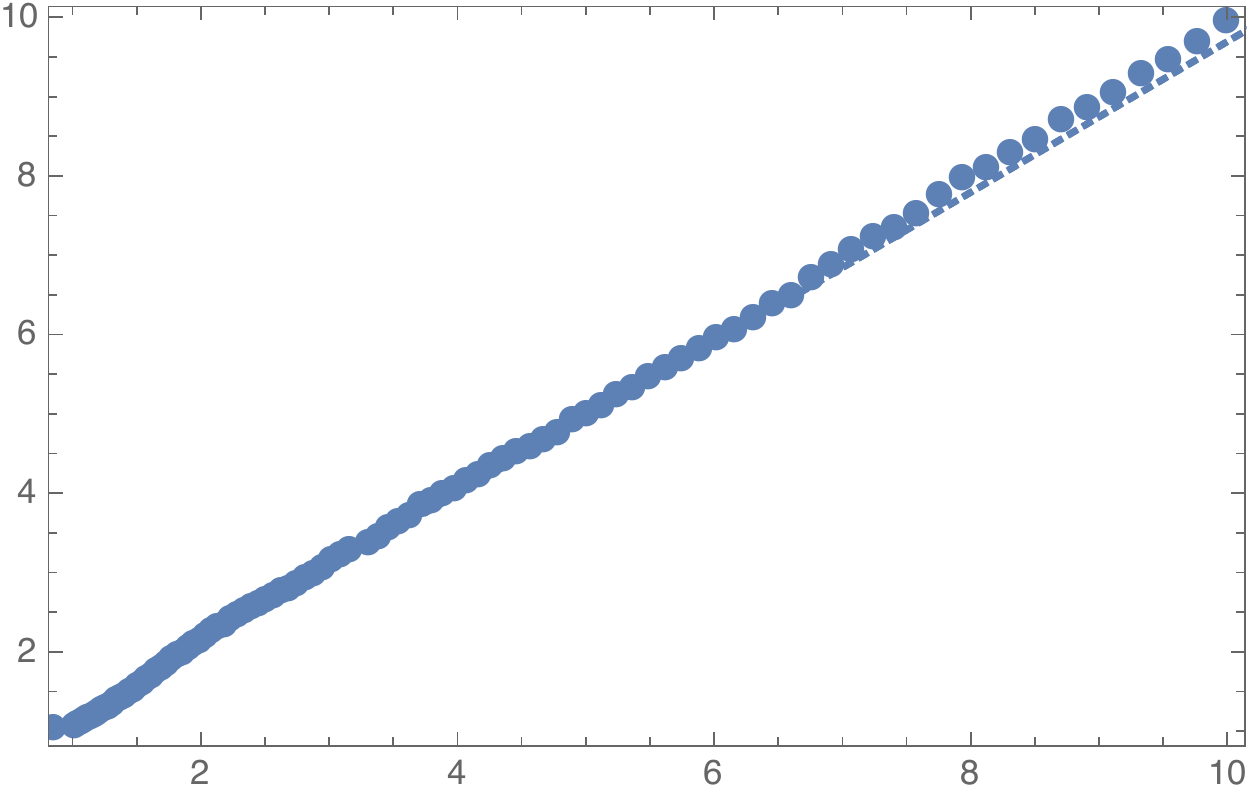}} \\
	\subfigure[Scubert]{
		\includegraphics[width=0.4\textwidth]{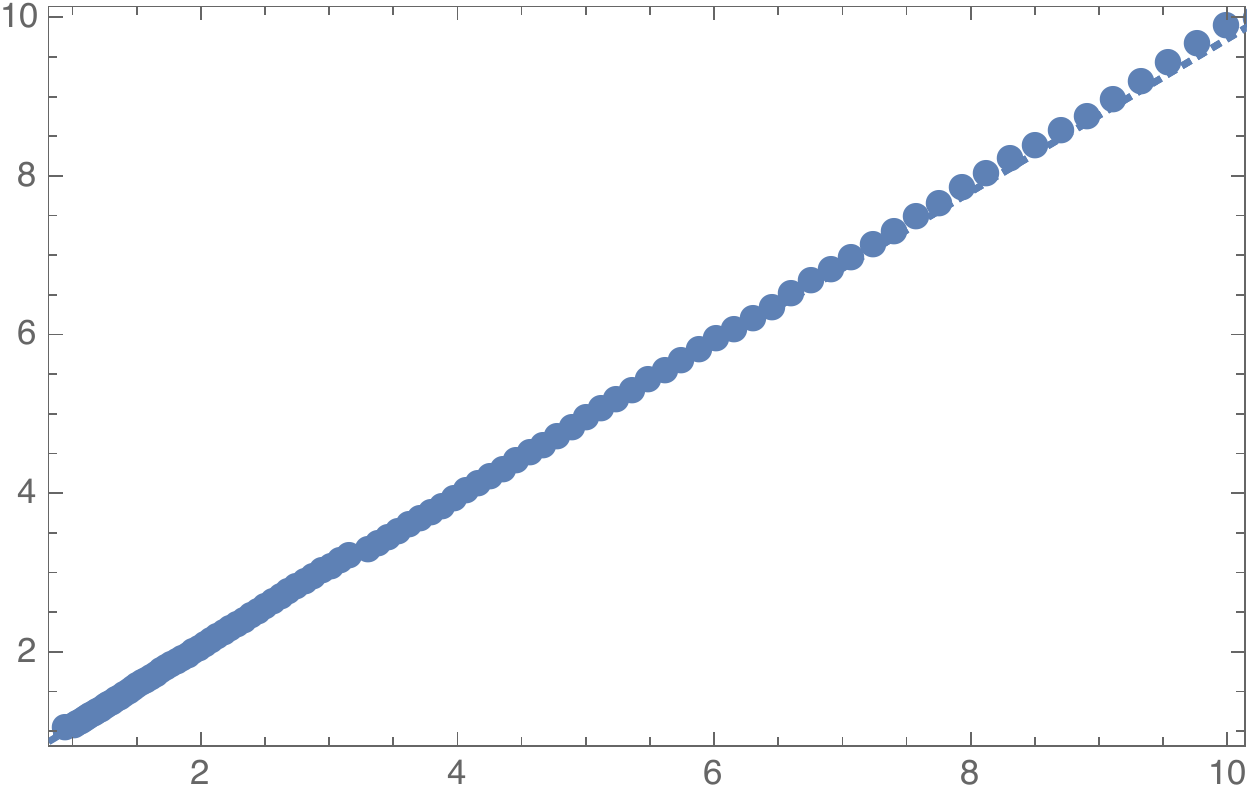}} 
	\quad
	\subfigure[All composers using 521 files]{
		\includegraphics[width=0.4\textwidth]{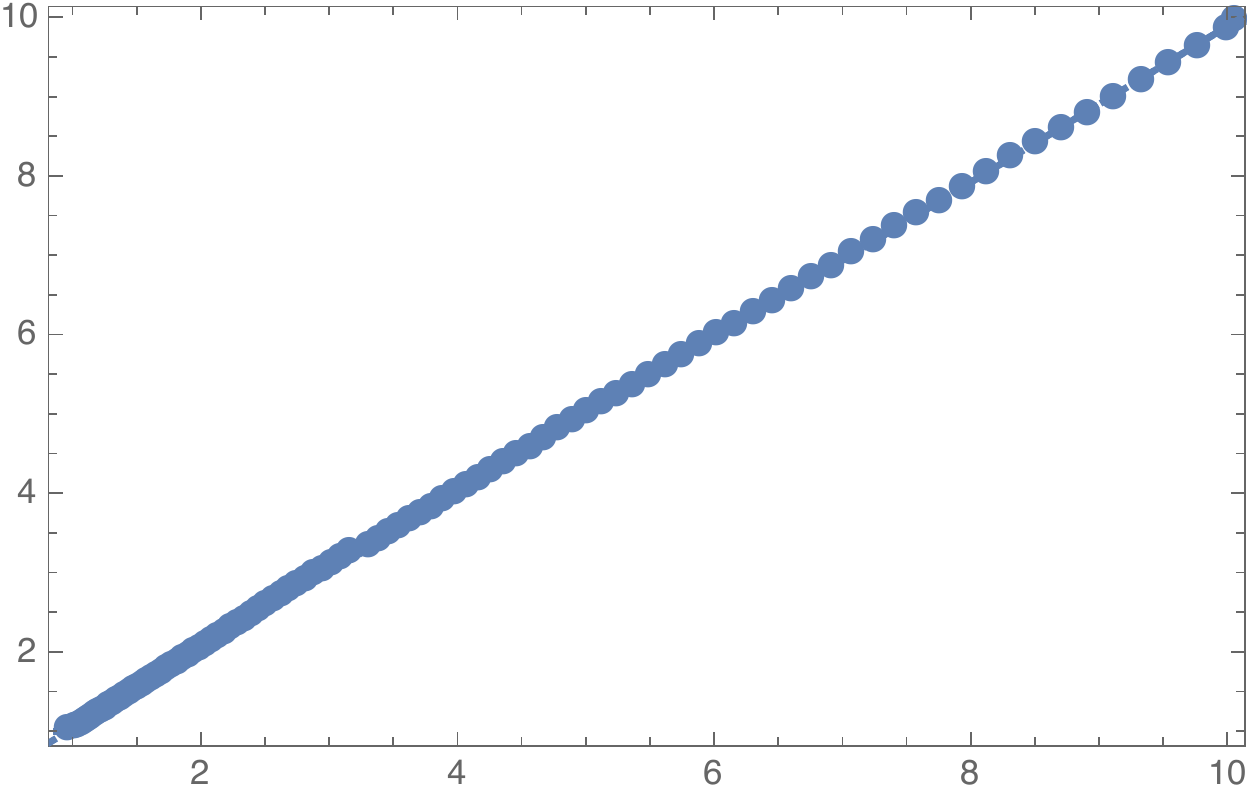}} 
	\caption{Quantile-Quantile Plots.}
\end{figure}

Motivated by these observations, we examined a large \footnote{The complete selection contains 72 pieces by Bach, 32 pieces by Beethoven, 41 pieces by Mozart, 271 pieces by Schubert, and 105 pieces by Tchaikovsky.} selection of music files by J. S. Bach, Mozart, and Schubert, and for each composer we found Benford distributed time durations. Finally, we took the union of all data sets, we obtained a close-to-perfect Benford conformance. The results are presented in Figure \ref{fig:all}.

In conclusion, based on our 
analysis,  we would like to advance the conjecture that the time durations in classical pieces are Benford distributed.

Preliminary work done on different genres of music such as Blues, Jazz, and Rock indicates that our observation applies beyond the classical music. We plan to present these results once completed.

\section*{Disclosure statement}%
\addcontentsline{toc}{section}{Disclosure statement}

No potential conflict of interest was reported by the authors. \\ 

\section*{Data availability statement}
\addcontentsline{toc}{section}{Data availability statement}%
The data that support the findings of this study are available from the corresponding author, AK, upon reasonable request.

\addcontentsline{toc}{section}{References}

\end{document}